# Bulk nanostructured AlCoCrFeMnNi chemically complex alloy synthesized by laser-powder bed fusion process

*Hyo Yun Jung*[1*], Nicolas J. Peter[1], Eric Gärtner[2], Gerhard Dehm[1], Volker Uhlenwinkel[2], and Eric A. Jägle[1]

[*]Corresponding Author: h.jung@mpie.de
[1.] Max-Planck-Institut für Eisenforschung GmbH, 40237 Düsseldorf, Germany
[2.] Leibniz-Institute for Materials Engineering IWT, 28359 Bremen, Germany

**Abstract**

We report the synthesis of a bulk nanostructured alloy using the laser-powder bed fusion process. The equiatomic AlCoCrFeMnNi chemically complex alloy forms a nanoscale modulated structure, which is homogeneously distributed in the as-built condition. The nanostructure consisted of Al & Ni-rich ordered (B2) and Cr & Fe-rich disordered (A2) BCC phases. The two phases form an interconnected phase-network with coherent interface boundaries. Atom-probe-tomography and aberration-corrected scanning transmission electron microscopy analysis of the spatial distribution of the modulated structure suggests the occurrence of nano-scale spinodal decomposition. These results introduce a direct synthesis of bulk nanostructured alloys with promising geometric flexibility.

# 1. Introduction

Nanostructured alloys with a nano-scale substructure (typically below 100 nm) have attracted increasing attention over the years [1]. The alloys have been developed to achieve superior properties compared to conventional alloys. Contributing factors to these properties include small grain size and high-volume fractions of grain boundaries or interfaces [2]. As the nanostructured alloys often possess enhanced mechanical properties combined with improved functional properties such as biomedical [3,4], hydrogen-storage [5], and magnetic properties [6,7], they become technologically important for advanced applications.

Since the first report of nanostructured alloys by Gleiter [8], a number of methods have been developed for the bulk synthesis of nanostructured alloys, including severe plastic deformation (SPD) [9], partial crystallization of amorphous materials [10], and consolidation of powders with nano-scale structure [11,12]. However, these processes still have limitations regarding the shape, size, and quantity of material that can be produced, restricting their otherwise extensive applications [13].

Additive manufacturing (AM) builds three-dimensional parts layer-by-layer, allowing the synthesis of bulk parts with high geometric flexibility [14]. Among various technologies, laser-powder bed fusion (L-PBF) is known to be suitable for synthesizing metallic materials with high relative density and complex geometry [15]. Additionally, due to the small melt pools generated in the process, it induces a high cooling rate ($10^4$~$10^6$ K/s) in the molten metal, which generally leads to the formation of a non-equilibrium microstructure [16]. A number of metallic materials that show a limited processability by conventional rapid cooling processes such as splat quenching and melt spinning, and are therefore limited in the geometrical size, have been successfully synthesized by the AM process. This includes bulk metallic glasses [17] and high-entropy alloys (HEAs) [18].

To date, there has been limited research on the direct synthesis of bulk nanostructured material using metal AM processes. There have been a number of studies on ceramic-reinforced metal matrix composites [19-21], but this type of particle reinforcement is not suitable for all applications. Moreover, like conventional composite synthesis, it requires an addition of nanoparticles to the metal matrix already during the preparation of the feedstock powder. Since the L-PBF process can directly synthesize metastable microstructures without such specialized feedstock preparation, it can provide a novel pathway to form bulk nanostructured materials with high flexibility in the geometry of the produced part. When the rapid cooling condition of the L-PBF process suppresses long-range atomic diffusion for phase coarsening, we can expect a direct formation of nanostructured alloys in the as-built condition.

In this work, we synthesize bulk nanostructured materials directly using the L-PBF process. We firstly look for an alloy system with a strong tendency to form a two-phase microstructure by phase separation. After that, we aim to refine the resulting microstructure up to the point of it becoming nano-scaled. As a suitable candidate material, we selected multi-component chemically complex alloys (CCA), an expanded concept of HEAs containing more than one phase [22]. While the original concept of HEAs follows the idea of forming a single FCC, BCC, or HCP-type solid solution [23], transition metal-based HEAs have been reported to have a tendency of phase separation [22,24]. Among them, the HEAs with Al addition have been extensively studied as CCAs owing to their interesting phase separation behavior as a function of Al contents [25-27]. Increasing the Al content in transition metal-based HEAs leads to a phase transition of the main phase of the alloy from FCC to BCC. Such Al containing alloys in a near-equiatomic concentration can form a duplex Fe-Cr-rich disordered BCC (A2, W-type structure) and Al-Ni-rich ordered BCC phase (B2, CsCl-type structure). While the phase separation mechanism of the Al-containing alloys is not well understood yet, it is speculated to be due to spinodal decomposition [28]. Tang et al. investigated the role of Al in the phase separation, revealing that the Al addition leads to strong covalent bonds between Al and the surrounding transition metal (TM) atoms and a stabilized ordered BCC phase structure [29]. Therefore, the Al-containing alloys in a near-equiatomic concentration form a duplex A2/B2 phase, probably through a spinodal decomposition mechanism.

In the case of the Al-containing CCAs, Karlsson et al. [30] recently reported the synthesis of the equiatomic AlCoCrFeNi alloy using the L-PBF process. While the atom probe analysis of the alloy showed nonperiodic fluctuations of the Cr concentration in a range of 17 to 40 at. %, in general, it formed a relatively homogeneous BCC-type structure without nano-scale elemental decomposition. Considering the fact that the phase separation by spinodal decomposition depends strongly on alloy composition, here, we attempted a compositional modification to promote the nano-scaled spinodal decomposition. We select Mn as an additional element to the AlCoCrFeNi alloy. According to Tang et al. [29] the Al in transition metal-based HEAs leads to the formation of closely coupled Al-TM clusters, as the Al atom has a large atomic radius and high negative heat of mixing with Ni (- 22 kJ/mol) and Co (- 19 kJ/mol). This reduces atomic diffusivity and promotes the spinodal decomposition in transition metal-based HEAs. Here, w can expect that the Mn may join Al-TM clusters, which further enhances the tendency of spinodal decomposition. As shown in Table 1, the Mn atom has a high negative heat of mixing with Al (- 19 kJ/mol). In addition, as compositional fluctuations during the L-PBF process can trigger spinodal decomposition, this alloying approach seems suitable to our purpose.

In this paper, we demonstrate the synthesis of bulk nanostructured alloys via the L-PBF process of the AlCoCrFeMnNi CCA. X-ray diffraction (XRD), scanning electron microscopy (SEM) and aberration-corrected scanning transmission electron microscopy (STEM) are performed to identify the crystal structures, microstructure and elemental distribution of the specimen. Additionally, we perform atom probe tomography (APT) to reveal the three-dimensional distribution and compositional fluctuations of the nano-scale structure. Based on these results, we introduce a new approach of bulk synthesis for nanostructured alloys with the combined advantages of the AM process and CCAs.

Table 1. The atomic radius, structure type, electronegativity and mixing enthalpies of atomic pairs between elements, ΔHmix (kJ/mol), calculated by Miedema's model [31].

|    | Atomic radius [nm] | Structure type | Al | Co  | Cr | Fe  | Ni  | Mn  |
|----|--------------------|----------------|----|-----|----|-----|-----|-----|
| Al | 0.143              | FCC (A1)       | *  | -18 | 0  | -11 | -22 | -19 |
| Co | 0.125              | HCP (A3)       |    | *   | -4 | -1  | 0   | -5  |
| Cr | 0.125              | BCC (A2)       |    |     | *  | -1  | -7  | 2   |
| Fe | 0.124              | BCC (A2)       |    |     |    | *   | -2  | 0   |
| Ni | 0.124              | FCC (A1)       |    |     |    |     | *   | -8  |
| Mn | 0.135              | αMn (A12)      |    |     |    |     |     | *   |

## 2. Experimental methods

A pre-alloyed ingot with a nominal composition of $Al_{16.66}Co_{16.66}Cr_{16.66}Fe_{16.67}Mn_{16.66}Ni_{16.66}$ (at.%) was prepared under Ar atmosphere by induction melting of pure elements (Al, Co, Cr, Fe, Mn, and Ni) with commercial-grade purity (> 99.9 %). Spherical powder with a particle size of less than 90 µm was prepared for the L-PBF process by inert gas atomization and subsequent sieving. The flowability of the powder was evaluated by the Hausner ratio [32] and angle of repose [33]. Bulk specimens with dimensions of 5 mm x 5 mm x 10 mm were fabricated by the L-PBF process on an austenitic stainless-steel substrate plate using an Aconity 3D Mini device equipped with a Yb-YAG laser under Ar gas atmosphere. The employed parameters of the L-PBF process are listed in Table 2.

Phase analysis of the bulk specimens was first performed by XRD in reflection geometry using a laboratory Philips PW 1830 diffractometer with Co radiation ($\lambda$ = 1.78897 Å). Diffractograms were acquired at a step size of 0.03°, with a count time of 20 s per step. The microstructure and elemental distribution of the as-built specimen were analyzed by backscatter electron (BSE) imaging and energy-dispersive X-ray spectroscopy (EDS) maps using a Zeiss Crossbeam FIB-SEM microscope equipped with TSL/EDAX detector (Digiview 1612 camera). The electron backscatter diffraction (EBSD) maps with an area of 400 x 800 µm² were collected with an acceleration voltage of 30 kV and a 80-nm step size. All measurements were performed on a cross section of the specimen containing the build direction. Tip-shaped specimens for atom probe tomography analysis were prepared with a ThermoFisher Helios Nanolab 600i FIB/SEM dual-beam device equipped with an Omni Probe micromanipulator. The final sharpening of the tips was carried out with an acceleration voltage of 5 kV and a current of 8 pA to minimize the Ga contamination on the surface due to FIB preparation. The samples were analyzed in the voltage pulsing mode of a Cameca LEAP 3000HR X local electrode atom probe instrument. At the operating temperature of 65 K, more than 10 million ions were detected per sample at a pulse frequency of 200 kHz and pulse fraction of 0.15. The software IVAS version 3.6.18 was used for reconstruction and data analysis. Wet chemical analysis of powder and as-built samples was performed by inductively coupled plasma optical emission spectrometry (ICP-OES). TEM lamellaes were prepared by a ThermoFisher Helios 600 crossbeam machine at 30 kV. Final polishing of the lamellae was performed at 5 kV. The samples were analyzed in an aberration-corrected ThermoFisher Titan Themis 60-300 microscope using an acceleration voltage of 300 kV and at semi-convergence and semi-collection angles of 23.8 mrad and 76-200 mrad, respectively. Micrographs were acquired in high-angle annular dark field (HAADF) conditions to make use of the associated Z-number contrast. The presented EDS map was collected with a super-X EDS system. The equilibrium phase diagram of the investigated alloy was calculated using ThermoCalc and the TCHEA3 database.

Table 2 Building parameters of the L-PBF process for the bulk synthesis of nanostructured AlCoCrFeMnNi alloy.

|  | Value |
|---|---|
| Atmosphere | Ar |
| Scanning strategy | 90° Alternating |
| Laser power [W] | 200 |
| Scan speed [m/s] | 0.50 |
| Layer thickness [μm] | 70 |
| Hatching distance [μm] | 90 |
| Beam diameter [μm] | 90 |
| Energy density [J/mm$^3$] | 127 |

## 3. Results

We first examined the effect of Mn addition on the phase formation of the alloy. Figure 1(a) shows a phase diagram of $(AlCoCrFeNi)_{1-x}Mn_x$ (x = 0 – 20 at.%) alloy calculated with Thermo-Calc software using the TCHEA3 database. As the Mn content increases, the formation of A1 phase is predicted to be suppressed, but the sigma phase in the low temperature range and coexistence of A2 and B2 phases in the high temperature range to be expanded; this implies an enhanced phase stability of the sigma, A2 and B2 phases with Mn addition. For the present study, we selected the specific Mn content of 16.6 at% to keep the alloy system simple with an equiatomic concentration of all the elements.

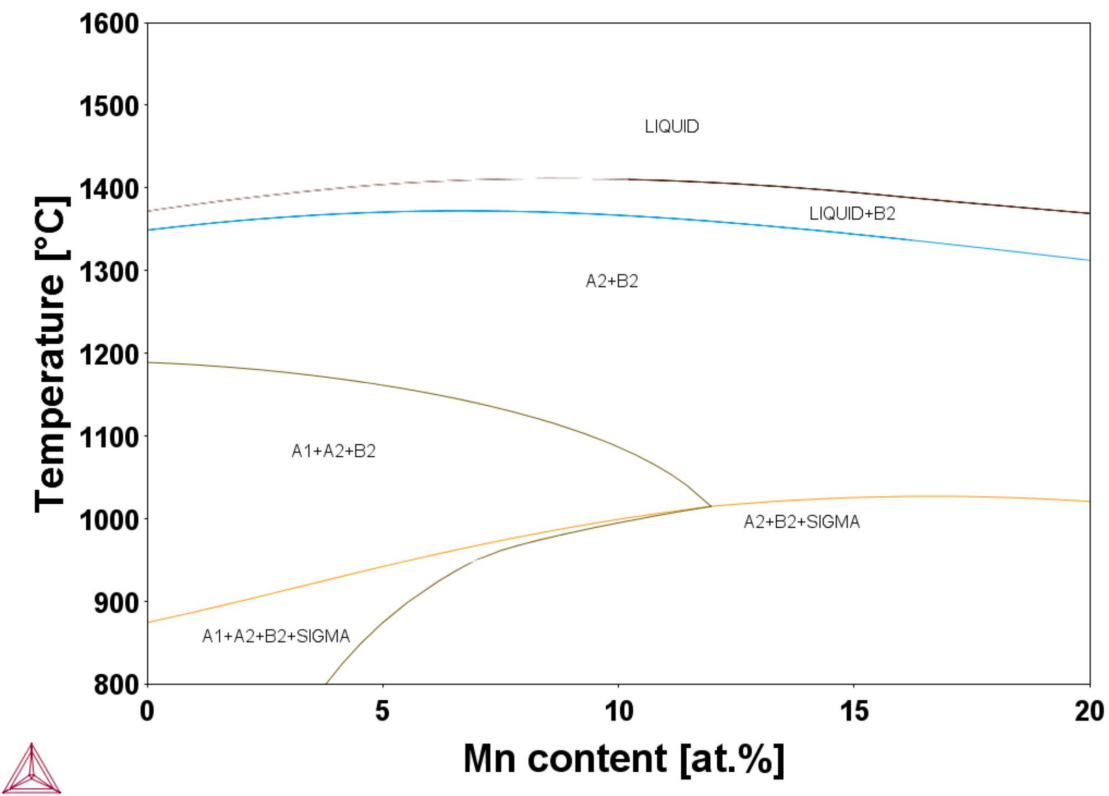

Figure 1 Predicted phase diagram of $(AlCoCrFeNi)_{100-X}Mn_x$ (x =0-20 at.%) alloy calculated by Thermo-Calc software.

Figure 2 (a) and (b) show the particle morphology and size distribution of the atomized feedstock powder for the L-PBF process. The basic characteristics of the powders are listed in Figure 2 (b). First, Figure 2 (a) shows that most particles have a rather smooth surface without satellites and a near-spherical shape, but that the fine particles tend to agglomerate. Considering the particle size distribution shown in Figure 2 (b), it is notable that the powder contained a relatively large number of fine particles, which usually deteriorates the flowability of the L-PBF process. However, the Hausner ratio of 1.138 and angle of repose of 32.1° listed in Figure 2 (b) reveal a sufficient flowability, while the powder does not flow through a Hall flowmeter at all [34]. Therefore, we could use this powder to build bulk specimens in rectangular shapes by L-PBF process (see Figure 2 (c)). The microstructure of the bulk specimen was initially investigated with cross-sectional backscatter electron (BSE) images. As shown in Figure 2 (d). It is notable that the AlCoCrFeMnNi alloy had a strong morphological texture composed of columnar grains approximately 30 μm in width. The coarse columnar grains were elongated in the building direction, which implies a high thermal gradient and rapid solidification rate during the applied process [35]. Microstructural defects, including micro-pores and cracks, were also visible in this specimen (Figure 2 (d)). Considering the size and morphology, the micro-pores with a spherical shape could be identified as entrapped gas pores [36].

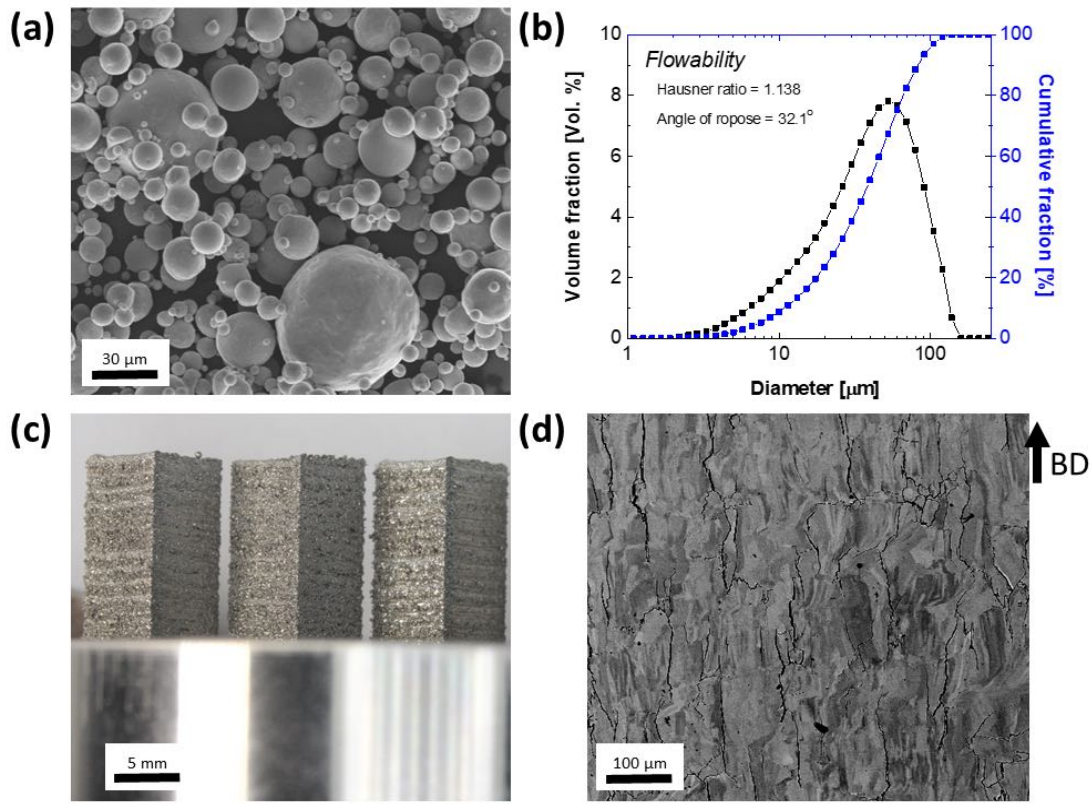

Figure 2 (a) SEM image and (b) particle-sized distribution of feedstock powder. (c) Image of as-built rectangular specimens and (d) a cross-sectional BSE image (build direction indicated). A preferred solidification direction with solidification crack is visible.

The phase content of the as-built AlCoCrFeMnNi alloy was investigated by XRD (Figure 3). Due to the strongly elongated grain structure observed in Figure 2 (d), both the horizontal section (XY) and vertical section (Z) of the bulk specimen were analyzed by XRD for texture and compared to a sample of the feedstock powder with a random orientation distribution. All diffraction patterns showed peaks of the primitive BCC phase (A2 phase) together with a super-lattice peak (at $2\theta = 36.29$ °), indicating the presence of ordering in BCC structure. By comparing the diffraction patterns of the bulk specimen with that of the parent powder, the bulk specimen examined with the vertical section (Z) showed considerably enhanced diffraction intensity at $2\theta = 36.29$ ° (100) and 77.10 ° (200), implying strong crystallographic textures with the {100} plane parallel to the building direction.

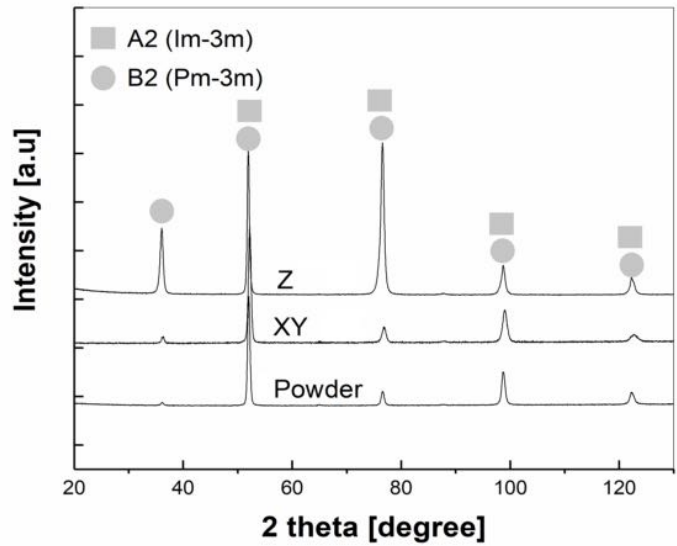

Figure 3 XRD patterns of parent powder and bulk specimen of AlCoCrFeMnNi alloy examined in vertical (Z) and transversal (XY) sections.

Elemental distribution near high-angle grain boundaries (with misorientation of over 15°) was investigated at the micro-scale by a combined analysis of EBSD and EDS mapping. Figure 4 (a) shows an inverse pole figure (IPF) map with high-angle grain boundaries marked as solid black lines. The corresponding EDS maps are shown in Figure 4 (b). While the Mn (and only the Mn) showed some inhomogeneity on the length scale of multiple grains (~100 μm), the EDS results reveal that the specimen had an otherwise homogeneous elemental distribution on the micro-scale, without apparent grain boundary segregation.

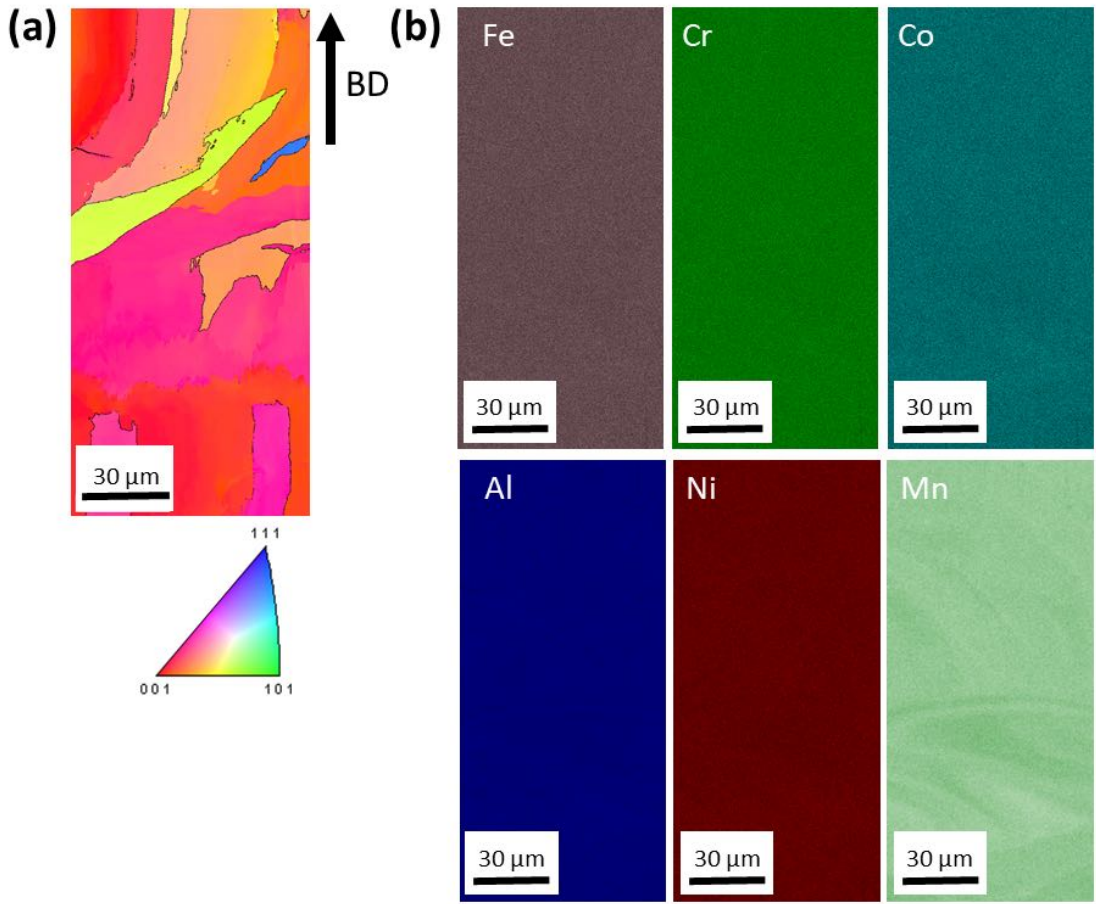

Figure 4 (a) Inverse pole figure and (b) EDS map of the bulk specimen of AlCoCrFeMnNi alloy. The black lines in (a) denote high-angle grain boundaries with misorientation angles over 15°.

The microstructure and elemental composition of the as-build bulk AlCoCrFeMnNi alloy were characterized using STEM imaging and APT at the nano-scale, respectively. In Figure 5 (a) and (b), the HAADF image shows a nano-scaled modulated structure with a coherent interphase boundary. The structure has an almost constant modulation width of approximately 10-15 nm. The EDS map shown as inset in Figure 5(b) indicates a clear elemental separation, i.e. blue shows the Ni and Al signal, while red is associated with Fe and Cr signals. The atomic resolution micrographs and Fast Fourier transform (FFT) patterns from the areas marked in the EDS map reveal the existence of the Fe/Cr-rich A2 and Ni/Al-rich B2 phases, due to the clear presence of the {001} super lattice reflections in (c) (highlighted by white circles) for the B2 phase compared to the A2 FFT pattern in (d). This separation is analyzed in detail by APT.

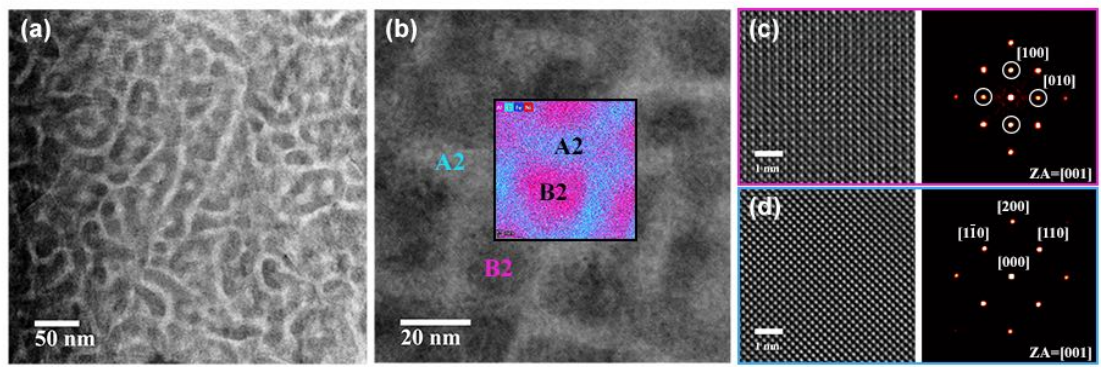

Figure 5 (a) STEM-HADDF overview image and (b) higher magnification image of the phase separated microstructure. The inset in (b) provides an EDS map of the exact location. (c) and (d) provide atomic resolution micrographs of the B2 and A2 phases highlighted in the EDS map, respectively, along with their corresponding Fast Fourier transforms.

Figure 6 (a) shows the three-dimensional volume reconstruction of a representative APT dataset. It is also notable that the elements are clearly decomposed. Figure 6 (b) shows the elemental distribution of a 4 nm thick vertical slice taken from the center of the APT dataset shown in Figure 6 (a), revealing the interconnected structure of Fe/Cr-rich and Ni/Al-rich regions. The nano-scaled structure is present throughout the entire three-dimensional volume of the APT tip. Figure 6 (c) shows the 1D composition profile along the region of interest marked with a cylinder in Figure 6 (a). Clear compositional oscillations of the elements can be identified. Fe and Cr co-partition, while the other elements (Al, Ni, Co, and Mn), show the opposite partitioning pattern. While the strongest compositional oscillation is found for Cr with an amplitude of approximately 40 at. %, Mn and Co concentrations also slightly rise where Ni and Al increased, implying a positional correlation of these elements. An unexpectedly low overall Mn content below 16.6 at% is also observed in Figure 6 (c).

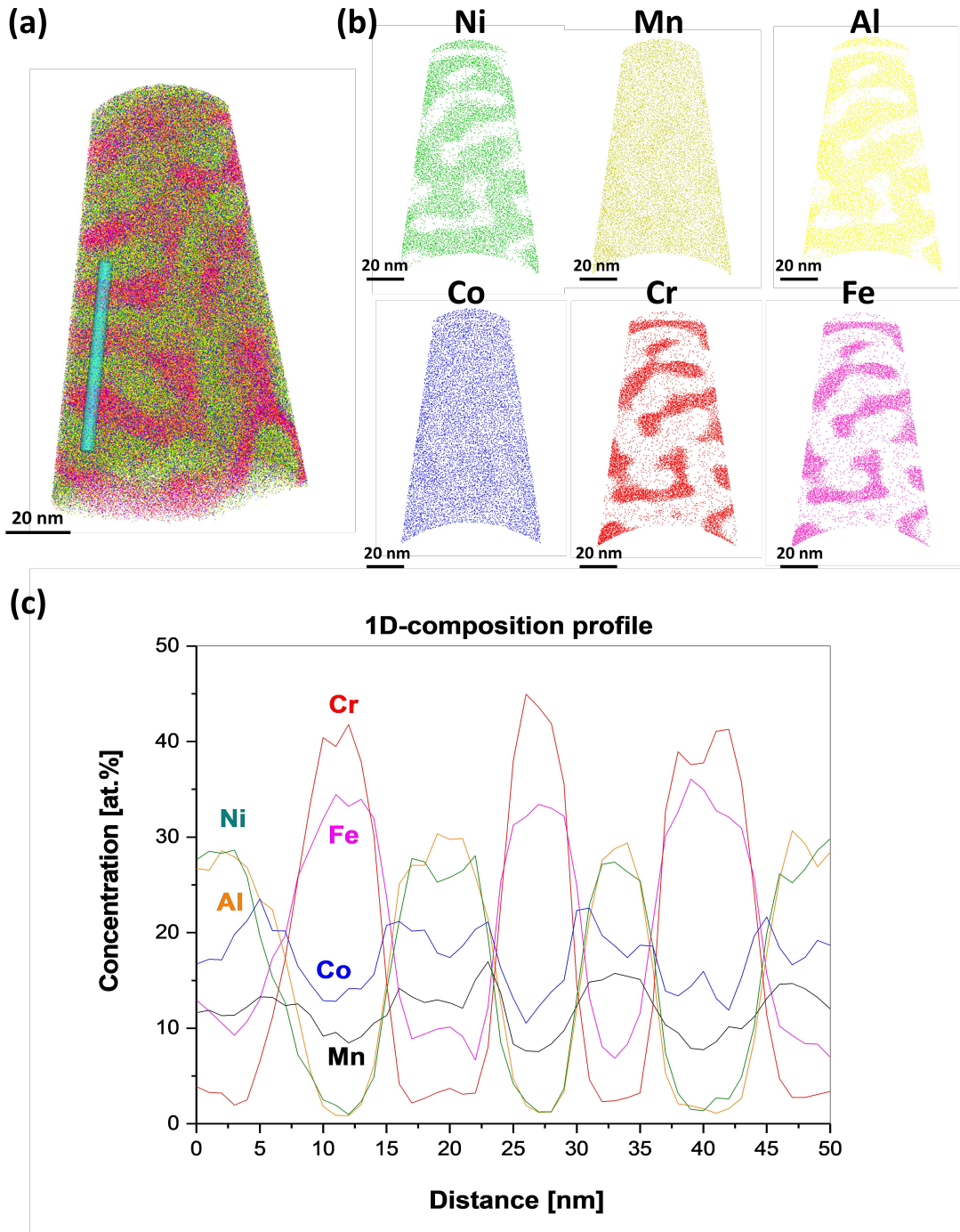

Figure 6 (a) Three-dimensional reconstruction of the APT data and its vertical sectional image of the elements, (b) elemental distribution within a slice taken from the center of (a) (thickness: 4 nm), and (c) 1D composition profile of the marked cylinder in (a).

# 4. Discussion

Here, we demonstrated the feasibility of the direct synthesis of nanostructured materials by the L-PBF process. We applied the L-PBF process to the Al-containing HEA with a strong phase separation tendency because it provides advantages of high dimensional flexibility and a rapid solidification rate. To promote the tendency of phase separation, Mn was added as the sixth element of the Al-containing HEA. After the L-PBF process, it was found that the bulk specimen in the as-built condition possessed a highly textured micro-scaled grain structure containing a nano-scaled interconnected two-phase.

It was observed that only Mn was distributed inhomogeneously at the micro-scale (Figure 4). This inhomogeneity is connected to the elemental characteristics and the process conditions. We interpret the distribution of the Mn-lean areas in a "flow-shaped pattern" across several grains as the result of partial Mn evaporation from the melt pool during the L-PBF process. Due to the high energy input induced by the focused laser beam, strong temperature gradients in the melt pool occur, and Marangoni convection results [37]. As high laser power and slow scanning speeds are used in this study to aid melting and to be able to process relatively coarse powder, the vapor pressure of Mn increased beyond the ambient pressure, aiding the convective flux and leading to the evaporation of Mn [37]. Note that Mn is the element in the investigated HEA that is most strongly impacted by evaporation, because it is the element with the lowest boiling point, cf. Figure 7 (a). To confirm the effect of the high laser energy on the Mn evaporation, the chemical compositions of the feedstock powder and bulk specimens synthesized with two different laser input energies were investigated through chemical analysis (Figure 7 (b)). The specimen with high laser input energy was prepared with half of the otherwise used scan speed ($v$ = 0.25 m/s), i.e. twice the laser input energy. As expected, the Mn contents of the bulk specimens were lower than those of the powders. Furthermore, the bulk specimens prepared with high laser input energy had a much lower Mn content than the ones with low laser input energy.

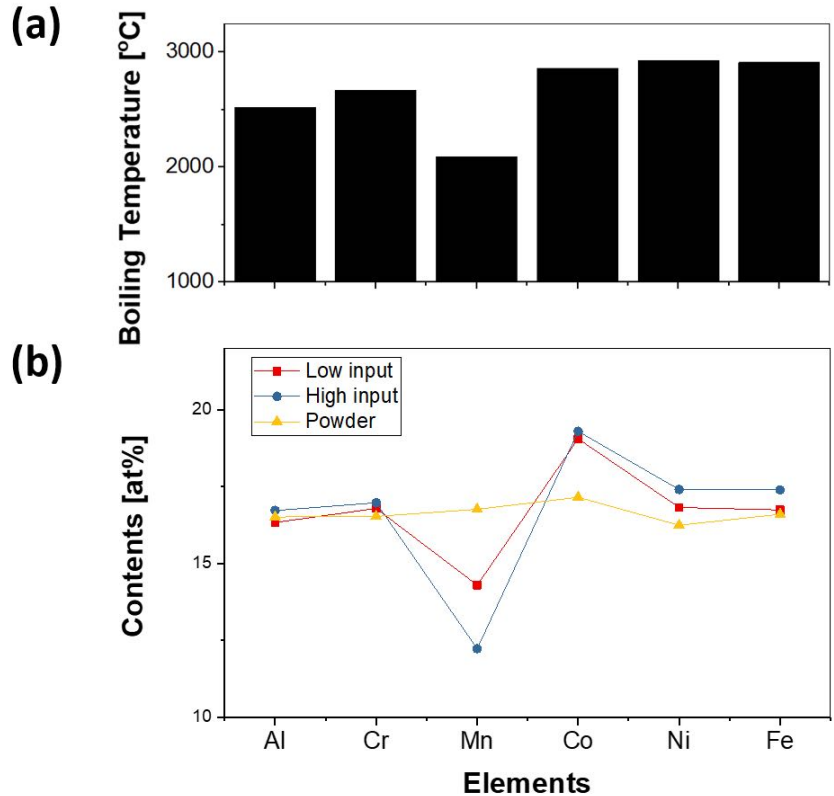

Figure 7 (a) Boiling temperature of the components and (b) chemical compositions of the powder, and bulk specimens synthesized with two-laser input energy.

We found that the alloy possessed a highly textured micro-scaled structure. The L-PBF process provides highly directional cooling conditions. The thermal gradient and solidification velocity at the solid–liquid interface affects the process of columnar grain and equiaxed grain formation [38]. The formation of columnar grains on the micro-scale is well known in cubic materials synthesized by L-PBF process [39]. Because <100> is the direction of fastest heat conduction in cubic structures, bulk specimens synthesized by the L-PBF process typically have a highly textured microstructure elongated in the build direction [34].

Throughout the elongated columnar grains, a modulated structure of A2 and B2 phase was homogeneously distributed. We characterized the spatial distribution and morphology of this modulated structure to understand the formation mechanism of the observed nanostructure. Through the APT results shown in Figure 6 (b), it was found that the fine A2 and B2 phases form an interconnected phase network with convex and concave curvature. The phase separation into A2 and B2 can be caused by two possible mechanisms, namely nucleation-and-growth or spinodal decomposition [40]. According to Vijayalakshmi et al., there are three stages of spinodal decomposition: initial spinodal stage, coarsening stage, and final particulate stage [41]. While the final microstructure of materials at the late stage of spinodal decomposition is difficult to distinguish from that formed by the nucleation and growth mechanism, the initial stage of spinodal decomposition shows a unique feature, an interconnected phase network caused by periodic concentration fluctuations. Therefore, in our case, it is highly likely that spinodal decomposition causes the nano-scale phase separation into A2 and B2 phases. On the other hand, it is notable that the distribution of the modulated structure is in an irregular pattern without preferential alignment along specific crystallographic directions, implying spinodal decomposition in the liquid phase [42]. While cubic alloys spinodally decomposed in the solid phase tend to propagate preferentially along elastically soft directions, like <100> [43,44], the spinodal decomposition in the liquid phase forms a continuous network structure with hyperbolic interfaces [45]. In our case, both the equilibrium liquid phase above the liquidus temperature and the metastable liquid phase undercooled below the liquidus temperature could be the origin of the spinodally decomposed structure. However, as the liquid-phase separation occurs usually in materials containing element pairs having a large positive enthalpy of mixing between elements [45], the present Al-containing HEA seems unlikely to show phase separation in the equilibrium liquid phase. As we applied the L-PBF process involving full melting and rapid solidification, a sudden undercooling of liquid phase can be easily obtainable. Our experimental results confirmed that only the interconnected A2 and B2 phases were formed in the as-built sample. Therefore, when a liquid phase was quenched into the A2 and B2 coexistence range, it becomes super-saturated and unstable. Then, density fluctuations are growing instantaneously, which leads to periodic concentration fluctuations and triggers spontaneous spinodal decomposition of undercooled liquid. In addition, the long-range atomic diffusion required for a phase coarsening might be effectively suppressed by the high solidification rate of the L-PBF process.

While we demonstrated the bulk synthesis of nanostructured alloys using AlCoCrFeMnNi HEA, this approach might be effective also for other metallic materials with a strong tendency to show spinodal decomposition. Spinodally structured alloys such as CuNiSn, AlNiCo or FeCrCo are usually synthesized by a complex process chain composed of casting, solution treatment, and spinodal treatment. However, the present approach can introduce a simple process with the additional advantage of high dimensional flexibility of the parts to be produced. While the present study demonstrated our first approach with BCC-type materials and the specimen contains a number of solidification cracks, the nano-scaled spinodal decomposition in HEAs may in principle provide a new opportunity to sustain high strength and extend ductility [29]. The basic approach of the present study actively uses the effect of rapid solidification rate on the formation of these unique micro- and nano-structures. Therefore, this approach is not limited to materials with spinodal decomposition, but can be extended to materials possessing strong phase separation tendency by the nucleation and growth mechanism.

## 5. Conclusions

The synthesis of a nano-scaled modulated A2/B2 structure was studied by processing the equiatomic AlCoCrFeMnNi High Entropy Alloy with the L-PBF process. The large thermal gradient and rapid solidification rate induced by the L-PBF process led to the formation of coarse columnar grains elongated in the building direction. While the high-energy laser beam of the L-PBF process caused some Mn inhomogeneity due to partial evaporation, the other elements were homogeneously distributed without any elemental segregation at the grain boundaries. The nano-scaled structure of the Fe/Cr-rich A2 phase and Ni/Al-rich B2 phase with an almost constant modulation width of approximately 10-15 nm was investigated with STEM and APT analysis and interpreted to be the result of liquid-phase spinodal decomposition from an undercooled melt. Considering the technological significance of the bulk nanostructured alloy, the present results suggest a new method for directly synthesizing a nano-scaled modulated structure.

## 6. Acknowledgements

Financial support from DFG for the "PaCCman" project in the framework of the Priority Programme "Compositionally Complex Alloys – High Entropy Alloys (CCA-HEA)" is acknowledged under the grant numbers DE 796/13-1, JA 2482/3-1, and UH 77/11-1.

## 6. References

[1] R.Z. Valiev, A.P. Zhilyaev, T.G. Langdon, Bulk Nanostructured Materials: Fundamentals and Applications, John Wiley & Sons, Inc, 2013.
[2] S. Cheng, J.A. Spencer, W.W. Milligan, Acta Mater. 51 (2003) 4505.
[3] R.Z. Valiev, I.P. Semenova, V.L. Latysh, H. Rack, T.C. Lowe, J. Petruzelka, L. Dluhos, D. Hrusak, and J. Sochova, Adv. Eng. Mater. 10 (2008) B15.
[4] I. Sabirov, N.A. Enikeev, M. Yu. Murashkin, R.Z. Valiev, Bulk nanostructured materials with multifunctional properties, Springer International Publishing, 2015.

[5] A. Zaluska, L. Zaluski, J.O. Ström–Olsen, J. Alloys Compd. 288 (1999) 217.
[6] G. Herzer, IEEE Trans. Magn, 26 (1990) 1397.
[7] D. Mishra, S. Sitaraman, S. Gandhi, S. Teng, P.M. Raj, H. Sharma, R. Tummala T.N. Arunagiri, Z. Dordi, R. Mullapudi, Electronic Components & Technology Conference (2015) 941.
[8] H. Gleiter, Prog Mater Sci. 33 (1989) 223.
[9] R.Z. Valiev, Nature Mater. 3 (2004) 511.
[10] K. Lu, Mater. Sci. Eng. R Rep. 16 (1996) 161.
[11] M.J. Mayo, Int. Mater. Rev. 41 (1996) 85.
[12] B.Q. Han, J. Ye, F. Tang, J. Schoenung, E. J. Lavernia, J. Mater. Sci. 42 (2007) 1660.
[13] M.J. Zehetbauer, Y.T. Zhu, Bulk Nanostructured Materials, Wiley-VCH, 2009.
[14] D.D. Gu, W. Meiners, K. Wissenbach, R. Poprawe, Int. Mater. Rev 57 (2002) 133.
[15] A.B. Spierings, M. Voegtlin, T. Bauer, K. Wegener, Prog Addit Manuf 1 (2016) 9.
[16] A. Mertens, S. Reginster, H. Paydas, Q. Contrepois, T. Dormal, O. Lemaire, J. Lecomte-Beckers, Powder Metall., 57(2014) 184.
[17] H.Y. Jung, S.J. Choi, K.G. Prashanth, M. Stoica, S. Scudino, S.H. Yi, U. Kuehn, D.H. Kim, K.B. Kim, J. Eckert, Mater. Des. 86 (2015) 703.
[18] Y. Brif, M. Thomas, I. Todd, Scr. Mater. 99 (2015)93.
[19] D. Gu, G. Meng, C. Li, W Meiners, R Poprawe, Scr. Mater 67 (2012) 185.
[20] J. Li, M. Su, X. Wang, Q. Liu, K. Liu, OPT LASER TECHNOL, 117 (2019) 158.
[21] D. Kong, C. Dong, X. Ni, L. Zhang, C. Man, G. Zhu, J. Yao, J. Yao, L. Wang, X. Cheng, X. Li, J. Alloys Compd, 803, 30 (2019) 637.
[22] A.M. Manzoni and U. Glatzel, Materials Characterization 147 (2019) 512.
[23] J.W. Yeh, S.K. Chen, J.Y. Gan, S.J. Lin, T.S. Chin, T.T. Shun, C.H. Tsau, and S.Y. Chang, Metal. Mater. Trans. A Phys. Metall. Mater. Sci., 35A (2004) 2533.
[24] L. J. Santodonato, P. K. Liaw, R. R. Unocic, H. Bei, and J. R. Morris, Nat Commun. 9 (2018) 4520
[25] J.W. Yeh, S.K. Chen, S.J. Lin, J.Y. Gan, T.S. Chin, T.T. Shun, C.H. Tsau, S.Y. Chang, Adv. Eng. Mater. 6 (2004) 99.
[26] X.D. Xu, P. Liu, S. Guo, A. Hirata, T. Fujita, T.G. Nieh, C.T. Liu, M.W. Chen, Acta Mater. 84 (2015) 145.
[27] U.S. Hsu, U.D. Hung, J.W. Yeh, S.K. Chen, Y.S. Huang, C.C. Yang, Mater. Sci. Eng. A 460-461 (2007) 403.
[28] A. Manzoni, H. Daoud, R. Volkl, U. Glatzel, N. Wanderka, Ultramicroscopy, 132 (2013) 212.
[29] Z. Tang, M.C. Gao, H. Diao, T. Yang, J. Liu, T. Zuo, Y. Zhang, Z. Lu, Y. Cheng, Y. Zhang, K.A. Dahmen, P.K. Liaw, T. Egami, JOM, 65(12), 1848.
[30] D. Karlsson, A. Marshal, F. Johansson, M. Schuisky, M. Sahlberg, J. M. Schneide, U. Jansson, J. Alloys Compd. 784 (2019) 195.
[31] A. Takeuchi, A. Inoue, Mater. Trans., 46 (2005) 2817.
[32] N. Harnby, A.E. Hawkins, D. Vandame, Chem Eng Sci. 42 (4) (1987) 879.
[33] D. Geldart, E.C. Abdullah, A. Hassanpour, L.C. Nwoke, I. Wouters, China Particuol 4 (2006) 104.
[34] A.B. Spierings, M. Voegtlin, T. Bauer, K. Wegener, Prog Addit Manuf 1 (2015) 1.
[35] M. Brandt, S. Sun, M. Leary, S. Feih, J. Elambasseril, Q. Liu, Adv. Mater. 633 (2013) 135.
[36] H. Gong, K. Rafi, H. Gu, T. Starr, B. Stucker, Addit. Manuf. 4 (2014) 87.

[37] V. Manakari, G. Parande, M. Gupta, Metals 7 (2017) 1.
[38] F. Yan, W. Xiong, E.J. Faierson, Materials 10 (2017) 1260.
[39] S. Bontha, N.W. Klingbeil, P.A. Kobryn, H.L. Fraser, Mater. Sci. Eng. A 513–514 (2009) 311.
[40] J.W.P. Schmelzer, A.S. Abyzov, J. Möller, J. Chem. Phys. 121 (2004) 6900.
[41] M. Vijayalakshmi, V. Seetharaman, V.S. Raghunathan, J. Mater. Sci. 17 (1982) 126.
[42] V. Sofonea, K.R. Mecke, Eur. Phys. J. B 8 (1999) 99.
[43] J.W. Cahn, Acta metal 9 (1961) 795.
[44] R. Zhao, J. Zhu, Y. Liu, Z. Lai, Rev. Adv. Mater. Sci. 33 (2013) 429.
[45] D.H. Kim, W.T. Kim, E.S. Park, N. Mattern, J. Eckert, PROG MATER SCI 58 (2013) 1103.